\documentclass{emulateapj}
\usepackage{epsf}
\usepackage{subeqn}

\slugcomment{submitted to ApJ}

\def\msun{\,{\rm M_\odot}}

\def\etal{{et al.\ }}
\def\ie{{\it i.e.\ }}
\def\spose#1{\hbox to 0pt{#1\hss}}
\def\lta{\mathrel{\spose{\lower 3pt\hbox{$\mathchar"218$}}
     \raise 2.0pt\hbox{$\mathchar"13C$}}}
\def\gta{\mathrel{\spose{\lower 3pt\hbox{$\mathchar"218$}}
     \raise 2.0pt\hbox{$\mathchar"13E$}}}

\def\max{{\rm max}}
\def\Om{\Omega}

\def\de{\delta}
\def\De{\Delta}
\def\s{\sigma}
\def\th{\theta}
\def\Md{{M_d}}
\def\Mc{{M_c}}
\def\lg{\log_{10}}
\def\heavi{{\cal H}}
\def\lcdm{$\Lambda$CDM\ }
\def\medc{\overline{c}_*}

\newcommand{\mincir}{\raise -2.truept\hbox{\rlap{\hbox{$\sim$}}\raise5.truept
\hbox{$<$}\ }}
\newcommand{\magcir}{\raise -2.truept\hbox{\rla669p{\hbox{$\sim$}}\raise5.truept
\hbox{$>$}\ }}
\newcommand{\minmag}{\raise-2.truept\hbox{\rlap{\hbox{$<$}}\raise 6.truept\hbox
{$>$}\ }}

\newcommand{\be}{\begin{equation}}
\newcommand{\ee}{\end{equation}}
\newcommand{\ba}{\begin{eqnarray}}
\newcommand{\ea}{\end{eqnarray}}
\newcommand{\brr}{\begin{array}}
\newcommand{\err}{\end{array}}
\newcommand{\bc}{\begin{center}}
\newcommand{\ec}{\end{center}}
\newcommand{\f}{\frac}

\newcommand\pvec{{\vec p}}
\newcommand\Dth{\Delta\theta}
\newcommand\sigc{\s_{\log c}}

\begin{document}

\lefthead{Kuhlen, Keeton, \& Madau}
\righthead{Lensing Statistics with Dark Energy}

\title{Gravitational Lensing Statistics in Universes Dominated by Dark Energy}

\author{Michael Kuhlen\altaffilmark{1}, Charles R. Keeton\altaffilmark{2,3},
\& Piero Madau\altaffilmark{1}}

\altaffiltext{1}{Department of Astronomy and Astrophysics, University of 
California, 1156 High Street, Santa Cruz, CA 95064.}
\altaffiltext{2}{Astronomy \& Astrophysics Department, University of
Chicago, 5640 S.\ Ellis Ave., Chicago, IL 60637.}
\altaffiltext{3}{Hubble Fellow.}

\begin{abstract}

The distribution of image separations in multiply-imaged gravitational
lens systems can simultaneously constrain the core structure of dark
matter halos and cosmological parameters.  We study lens statistics
in flat, low-density universes with different equations of state
$w=p_Q/\rho_Q$ for the dark energy component.  The fact that dark
energy modifies the distance-redshift relation and the mass function
of dark matter halos leads to changes in the lensing optical depth as
a function of image separation $\Dth$.  Those effects must, however,
be distinguished from effects associated with the structure of dark
matter halos.  Baryonic cooling causes galaxy-mass halos to have
different central density profiles than group- and cluster-mass halos,
which causes the distribution of normal arcsecond-scale lenses to differ
from the distribution of ``wide-separation'' ($\Dth \gtrsim 4\arcsec$)
lenses.  Fortunately, the various parameters related to cosmology and
halo structure have very different effects on the overall image
separation distribution:
(1) the abundance of wide-separation lenses is exremely sensitive
(by orders of magnitude) to the distribution of ``concentration''
parameters for massive halos modeled with the Navarro-Frenk-White
profile;
(2) the transition between normal and wide-separation lenses depends
mainly on the mass scale where baryonic cooling ceases to be efficient;
and (3) dark energy has effects at all image separation scales.
While current lens samples cannot usefully constrain all of the
parameters, ongoing and future imaging surveys should discover
hundreds or thousands of lenses and make it possible to disentangle
the various effects and constrain all of the parameters simultaneously.
Incidentally, we mention that for the sake of discovering lensed
quasars, survey area is more valuable than depth.

\end{abstract}


\section{Introduction}

Cold dark matter (CDM) theory makes robust predictions on the number
density, spatial distribution, and structural properties of dark matter
halos that must be compared with observational data to test the CDM
paradigm.  Given a large well-defined sample of strong gravitational
lens systems, the distribution of image separations $\Dth$ is a
powerful and direct probe of the halo mass function and inner density
profiles.  This probe is attractive for being independent of the
uncertainties about the relation between mass and luminosity that
plague most astrophysical tools.  As an example, the statistics of
wide-separation ($\Dth \gtrsim 6\arcsec$) lenses constrain the core
mass fraction of dark matter halos on group and cluster mass scales,
which depends on the ``concentration'' and slope of the central
density cusp (Keeton \& Madau 2001; also see Flores \& Primack 1996),
both of which are still uncertain and controversial (Navarro, Frenk,
\& White 1997; Moore \etal 1999; Jing \& Suto 2000; Ghigna \etal
2000).  On smaller, galaxy mass scales (corresponding to $\Dth$ of a
few arcseconds), the test is complicated by the presence of cooled
baryons; when baryons cool and condense into a galaxy they modify
the surrounding dark matter halo (e.g., Blumenthal \etal 1986).

The statistics of gravitational lensing are also sensitive to
cosmological parameters, since these determine the angular diameter
distances to the lens and the source, and the number density of lens
galaxies.  Observations of distant Type Ia supernovae (Riess \etal
2001; Perlmutter \etal 1999), combined with measurements of cosmic
microwave background (CMB) anisotropies (e.g., Spergel \etal 2003;
de Bernardis \etal 2002; Pryke \etal 2002; Balbi \etal 2000), provide
strong evidence that the dominant component in the universe --- the
exotic dark energy --- is not associated with matter, has negative
pressure, and is causing the cosmic expansion to accelerate.  While
several independent techniques appear to have converged rather tightly
on a ``concordance'' model with parameters $\Omega_{\rm tot}=1$,
$\Omega_M=0.3$, $h=0.7$, and $n=1$, determining the equation of state
of the dark energy remains one of the greatest challenges in cosmology
and physics today.  This may prove difficult with supernovae data alone
(Gerke \& Efstathiou 2002; Efstathiou 1999; Perlmutter, Turner, \&
White 1999), and additional observations (like CMB anisotropies, see
e.g.\ Frieman \etal 2002) may be required to determine the nature of
the negative pressure component.

Gravitational lensing statistics have already been used as another
probe of the cosmic equation of state (Cooray \& Huterer 1999; Waga
\& Friemann 2000; Sarbu, Rusin, \& Ma 2001; Chae \etal 2002; Huterer
\& Ma 2003).  Dark energy modifies the background cosmological line
element, which affects the lensing geometry.  It also modifies the
power spectrum of density fluctuations on large scales (Ma \etal 1999),
the rate of structure growth, the critical overdensity for spherical
top-hat collapse, and the overdensity at virialization (Wang \&
Steinhardt 1998; Weinberg \& Kamionkowski 2002), all of which affect
the mass function and the internal structure of collapsed dark matter
halos, with consequences for the lensing cross section and the
probability for multiple images.  Because dark energy varies with
redshift more slowly than matter, it starts contributing significantly
to the expansion only relatively recently, at $z \lesssim 1$, where
the lensing optical depth to distant quasars actually peaks.

In this paper we explore the ability of lens statistics to
simultaneously constrain the core structure of dark matter halos,
including both the concentration and the cooling mass scale, as well
as the halo mass function and the cosmic equation of state.  While
other recent studies have examined various aspects of the problem,
we consider all of the effects simultaneously to examine whether
lensing can really distinguish between them, and whether lensed
quasars can effectively be used to draw inferences on the background
cosmology.  The outline of the paper is as follows.  We first present
the ingredients of our calculations: the formalism for lens statistics
(\S\ref{sec:LensMeth}), and a description of structure formation
under the influence of dark energy (\S\ref{sec:StrForm}).  We then
study how the various parameters in our model affect the results
(\S\ref{sec:ParamDep}).  Next, we show that current lens data can
test the model and constrain some of the parameters
(\S\ref{sec:CLASScomp}).  Finally, we argue that ongoing and future
surveys should dramatically increase the samples of known lenses
and provide powerful constraints on the parameters relating to
dark energy and to the core structure of dark matter halos
(\S\ref{sec:Forecasts}).  We offer our conclusions in
\S\ref{sec:Conclusions}.

\section{Lensing Methods}
\label{sec:LensMeth}

In this section we present the formalism for lens statistics
calculations.  In \S\ref{sec:ProbCalc} we discuss the calculation
of various lensing probability distributions.  In
\S\ref{sec:Overview} we describe our model for the internal
structure of dark matter halos, in which halos corresponding to normal
galaxies are treated as singular isothermal spheres (SIS), while lower
and higher mass halos are assumed to have the Navarro-Frenk-White
(1997, hereafter NFW) profile.  In
\S\S\ref{sec:SISlens}--\ref{sec:NFWlens} we review the lensing
properties of SIS and NFW halos.

\subsection{Probability calculations}
\label{sec:ProbCalc}

We follow standard methods for computing lensing statistics (e.g.,
Narayan \& White 1988; Kochanek 1995; Porciani \& Madau 2000; Keeton
\& Madau 2001; Li \& Ostriker 2002).  Assume that the geometry of
the universe is well approximated on large scales by the
Friedmann-Robertson-Walker metric.  Consider a population of objects
that can act as deflectors, which lie at redshift $z_l$, have mass
$M$, and may be characterized by a set of additional parameters
$\pvec$ (such as a ``concentration'' parameter as defined in
\S\ref{sec:NFWlens}).  Let $dn/dM\,d\pvec$ be the differential comoving
number density of deflectors with the specified properties.  The
differential probability that a point source at redshift $z_s$ with
flux $S$ is lensed into multiple images by the deflector population
is then (e.g., Turner, Ostriker, \& Gott 1984)
\begin{eqnarray} \label{eq:dP}
  \frac{dP}{dz\,dM\,d\pvec} &=& (1+z_l)^3\,\frac{c\,dt}{dz}\,
    \frac{dn}{dM\,d\pvec} \\
  &&\quad\times \sigma_L(z_s,z_l,M,\pvec)\,B(S;z_s,z_l,M,\pvec) \nonumber
\end{eqnarray}
where
\begin{equation}
  \frac{c\,dt}{dz} = \frac{c}{(1+z)H(z)}
\end{equation}
is the cosmological line element, where $H \equiv \dot a/a$ is the
Hubble parameter and $a=(1+z)^{-1}$ is the scale factor.  In
eq.~(\ref{eq:dP}), $\sigma_L(z_s,z_l,M,\pvec)$ is the cross section
for multiple imaging, and $B(S;z_s,z_l,M,\pvec)$ is the ``magnification
bias'' factor representing the fact that magnification causes lensed
objects to be over-represented in flux-limited surveys.

The total lensing probability $P(z_s)$ can be computed by integrating
eq.~(\ref{eq:dP}) over redshift, mass, and the additional deflector
parameters $\pvec$.  The distribution of lensing observables, such as
the separation between the images, can be determined by inserting a
selection function in the integral.  For example, the probability of
lensing with an image separation greater than $\Dth$ is
\begin{equation}
  P(>\!\Dth;z_s) = \int dz_l\,dM\,d\pvec\
    \frac{dP}{dz_l\,dM\,d\pvec}\ \heavi(\Dth;z_s,z_l,M,\pvec)\,,
\end{equation}
where $\heavi(\Dth;z_s,z_l,M,\pvec)$ is unity if the parameters
correspond to a lens with image separation greater than $\Dth$, and
zero otherwise.  The probability of lensing with particular values
of the time delay or the lens redshift can be computed in the
analogous way.

We have computed the integrals with direct integration, and also with
Monte Carlo methods where we sum the integrand for random values of
the integration variables.  The advantage of the Monte Carlo approach
is the ability to compute the distributions of image separations,
time delays, and lens redshifts simultaneously, but the disadvantage
is the need for large-number realizations to reduce the statistical
fluctuations in the results.  We have verified that the two approaches
produce equivalent results.

\subsection{Overview of the model}
\label{sec:Overview}

The lensing cross section and magnification bias in eq.~(\ref{eq:dP})
depend on the radial density profile of the deflector.  Two standard
models for the density profile are the singular isothermal sphere
(SIS) and the Navarro-Frenk-White (1997, hereafter NFW) profile.
We cannot simply choose one or the other, however.  Assuming that all
massive objects have the same profile (be it SIS or NFW) produces a
distribution of lensed image separations that is grossly inconsistent
with the data at $\Dth \sim 1\arcsec$--$10\arcsec$ (Flores \& Primack
1996; Keeton 1998; Porciani \& Madau 2000; Li \& Ostriker 2002).  The
simplest model that agrees with the data has all objects below some
mass $M_{\rm cool}$ assumed to be SIS, and all halos above $M_{\rm cool}$
assumed to be NFW.  The transition from SIS to NFW is not ad hoc,
but rather motivated by baryonic cooling.  In large halos the baryons
have not had time to cool so the systems retain their initial NFW
form; while in small halos the gas has been able to cool, sink to
the center of the halo, and create a more concentrated profile that
can be approximated as SIS (Kochanek \& White 2001).  The lens data
require a mass threshold $M_{\rm cool} \sim 10^{13}\,h^{-1}\,M_\odot$,
which is consistent with the halo mass whose cooling time equals the
age of the universe, and also represents the mass scale of the
transition from galaxies to groups and clusters.

Recently Li \& Ostriker (2002) and Ma (2003) pointed out that similar
arguments also apply on the small-separation and low-mass end: there
must be a transition from SIS back to NFW (or a related form) as the
mass is decreased from normal galaxies ($\sim\!10^{12}\,M_\odot$) down
to dwarfs.  This transition reflects the possibility that feedback
effects or reionization could lead to a suppression of baryonic
collapse in low-mass halos (Springel \& Hernquist 2003; Bullock, Kravtsov, \& Weinberg 2002)

We adopt the two-transition-mass approach, where halos below mass
$M_d$ (``dwarfs'') have NFW profiles, halos between $M_d$ and $M_c$
have SIS profiles, and halos above $M_c$ (``clusters'') have NFW
profiles.  Some studies (e.g., Li \& Ostriker 2002) have used
generalized NFW profiles (Zhao 1996), where the central cusp is
steeper than for NFW, for dwarfs.  We always use the NFW form for
simplicity, but our results should apply to models with steeper
cusps as long as they have the same value of the core mass fraction
(Keeton \& Madau 2001).  We consider only spherical halos because
they are adequate for predicting the number of lenses (e.g., Keeton,
Kochanek, \& Seljak 1997; Chae 2002).

\subsection{SIS lens}
\label{sec:SISlens}

There is now strong evidence that the mass distributions of early-type
galaxies within several optical radii are well described by the SIS
profile; the evidence comes from lensing (Cohn \etal 2001; Rusin,
Kochanek, \& Keeton 2003), X-rays (Fabbiano 1989), stellar dynamics
(Rix \etal 1997), and even joint lensing and stellar dynamical
analyses (Koopmans \& Treu 2003; Treu \& Koopmans 2003).  An SIS
halo has density profile $\rho=\s_v^2/(2\pi G r^2)$, where $\s_v$ is
the velocity dispersion, which is related to the total mass $M$ within
the virial radius by
\begin{equation}
  \s_v(M,z) = \frac{1}{2}\,H_0^2\,r_0\,\Om_M^{1/2}\,\De_{\rm vir}^{1/6}\,
    (1+z)^{1/2}\ . \label{eq:sigma_vir}
\end{equation}
Here $r_0=(3 M/4 \pi \rho_0)^{1/3}$ is the comoving radius of the
collapsing perturbation, $z$ denotes the virialization epoch of the
halo, and $\De_{\rm vir}(z)$ is the virial overdensity of the halo
(see \S\ref{sec:tophat}).  SIS models are fully described by the mass
and redshift, so there are no additional parameters $\pvec$ and the
factor $dn/dM\,d\pvec$ in eq.~(\ref{eq:dP}) is just the mass function
$dn/dM$.

An SIS lens has an angular Einstein radius
\begin{equation}
  \th_E = 4\pi \left(\frac{\sigma_v}{c}\right)^2
    \frac{D_{ls}}{D_{os}}\ ,
\end{equation}
where $D_{ls}$ and $D_{os}$ are angular diameter distances from the
lens to the source and observer to the source, respectively.  The
image separation is $\Dth=2\th_E$, and the lensing cross section is
$\s^*_{\rm SIS} = \pi (\th_E D_{ol})^2$ where $D_{ol}$ is the angular
diameter distance from the observer to the lens.  The magnification
bias is
\begin{equation} \label{eq:bias1}
  B(S) = \frac{2}{\th_E^2} \int_{0}^{\th_E} dy\,y\,
    \frac{N(>\!S/\mu(y))}{N(>\!S)}\ ,
\end{equation}
where $N(>\!S)$ is the number of sources with flux greater than $S$.
The magnification $\mu(y)$ can be the magnification of either of the
two images ($\mu_\pm=\theta_E/y\pm1$) or the total magnification
($\mu_{\rm tot} = 2\theta_E/y$), depending on the details of a lens
survey.  For example, when studying the CLASS survey
(\S\ref{sec:CLASScomp}) we use the total magnification because the
targets in the initial catalog were unresolved.  Combining this with
a source luminosity function modeled as a power law
$dN/dS \propto S^{\nu}$ would yield a magnification bias of
$B = 2^{-\nu}/(3+\nu)$.

\subsection{NFW lens}
\label{sec:NFWlens}

High-resolution $N$-body simulations of the formation of dark matter
halos consistently produce halo density profiles shallower that SIS in
the core (NFW; Moore \etal 1999).  NFW argue that the halos form a
two-parameter family described by the density profile \be
\rho(r)=\f{\rho_s}{(r/r_s)(1+r/r_s)^2}\ , \ee where $\rho_s$ is a
characteristic density and $r_s$ is a scale radius.  In many cases it
is more convenient to take as the two free parameters the virial mass
$M$ and a ``concentration'' parameter $c=r_{\rm vir}/r_s$.  Note that
some authors define the concentration as $c=r_{200}/r_s$, where
$r_{200}$ is the radius at which the mean density of the halo is equal
to $200$ times the critical density (NFW; Li \& Ostriker 2002). This
definition is independent of cosmology, whereas the definition in
terms of the virial radius takes into account differences in the
virial overdensity for different cosmologies (see \S\ref{sec:tophat})

The two parameters are not actually independent; Bullock \etal (2001)
found that the concentration parameter follows a log-normal distribution
where the median depends on the halo mass and redshift,
\be 
  c_{\rm med}(M,z) = \frac{\medc}{1+z} \left(\frac{M}{M_*}\right)^{\alpha} ,
\ee
where $M_*$ is the mass of a typical halo collapsing today (see
\S\ref{sec:sigmaM}).  Thus, for NFW halos we can write the factor
$dn/dM\,d\pvec$ in eq.~(\ref{eq:dP}) as
\begin{equation}
  \frac{dn}{dM\,d\log c} = \frac{dn}{dM}\,\frac{1}{\sqrt{2\pi}\,\sigc}\,
    \exp\left[-\frac{(\log c - \log c_{\rm med})^2}{2\sigc^2} \right] .
\end{equation}
The halos in the \lcdm simulations by Bullock \etal were best
described by $\alpha=-0.13$ and $\medc=9.0$, with a scatter around the
median of $\sigc=0.14$~dex. (Note that Bullock \etal (2001)
incorrectly quote $\sigc=0.18$~dex. This error has been corrected in
Wechsler et al. (2002).) We find that at large separations the
abundance of lenses is very sensitive to both $\medc$ and $\sigc$, so
we treat both of them as model parameters. Other models for the
concentration distribution have been discussed, but they generally agree
out to $z \sim 1$ (Eke, Navarro, \& Steinmetz 2001; Wechsler et al.\ 2002;
Zhao et al.\ 2003a, 2003b).

Bartelmann (1996) derives the lensing properties of NFW halos, and
Oguri \etal (2002) give useful analytic approximations for the image
separation and magnification.  The lens equation has the form
\be \label{eq:NFWlens}
  y=x-\mu_s \f{g(x)}{x},
\ee
where $x$ is the magnitude of the position vector in the lens plane,
scaled by $r_s$, and $y$ the magnitude of the position vector in the
source plane, scaled by $r_s D_{ol}/D_{os}$.  The lensing efficiency
is parameterized by $\mu_s=4\rho_s r_s/\Sigma_{\rm cr}$, where
\be
  \Sigma_{\rm cr} = \frac{c^2}{4\pi G}\,\frac{D_{os}}{D_{ol} D_{ls}}
\ee
is the critical surface mass density for lensing (Turner \etal 1984).
Lastly,
\be
  g(x)\equiv 2\int_0^x \f{\Sigma(x')}{\Sigma_{cr}} x' dx',
\ee
which for the NFW profile evaluates to:
\be
  g(x)=\ln \f{x}{2} + \cases{
    \f{1}{\sqrt{1-x^2}} \mbox{arctanh} \sqrt{1-x^2}, & $0<x<1$, \cr
    1, & $x=1$, \cr
    \f{1}{\sqrt{x^2-1}} \arctan \sqrt{x^2-1}, & $x>1$.
}
\ee

The Einstein radius $x_E$ of the lens is the root of $y(x)$.  The
``radial critical radius'' $x_{\rm cr}$ is the root of $dy/dx$.
It maps to the radial caustic $y_{\rm cr}=|y(x_{\rm cr})|$.  Any
source with $|y| \le y_{\rm cr}$ is strongly lensed into three
images with $|x_1|>|x_2|>|x_3|$.  The image separation is
$\Dth=(|x_1|+|x_2|)r_s/D_{ol}$, so it depends on the source position
$|y|$.  However, Oguri \etal (2002) and Li \& Ostriker (2002) argue
that the separation is well approximated by
$\Dth \approx 2 x_E r_s/D_{ol}$.  We find that our results are
indistinguishable when we use this approximation and when we use the
exact image separation found by explicitly solving the lens equation.
The lensing cross section is $\s^*_{\rm NFW} = \pi(y_{\rm cr} r_s)^2$,
and the magnification bias is
\begin{equation} \label{eq:bias2}
  B(S) = \frac{2}{y_{\rm cr}^2} \int_{0}^{y_{\rm cr}} dy\,y\,
    \frac{N(>\!S/\mu(y))}{N(>\!S)}\ ,
\end{equation}
where $N(>\!S)$ is the number of sources with flux greater than $S$,
and $\mu(y)$ is the magnification of a source at impact parameter
$y$.  As in \S\ref{sec:SISlens}, the magnification may be that of
either image or the total magnification, depending on the details
of a lens survey.  The magnification must be found by explicitly
solving the lens equation and then using the fact that an image at
position $x$ has magnification $\mu = [y(x)/x]^{-1}[dy/dx]^{-1}$.

\section{Structure Formation in QCDM}
\label{sec:StrForm}

In this section we summarize the main features of structure formation
in a universe filled with dark matter and dark energy.  For
calculations of lens statistics, the quantities we need are the
geometry of the universe (\S\ref{sec:cosmography}), the mass function of
halos (\S\ref{sec:massf}) and the virial overdensity (\S\ref{sec:tophat}).

While the nature of the dark energy remains unknown, a cosmological
constant $\Lambda$ or a dynamical scalar field called ``quintessence''
(Caldwell, Dave, and Steinhardt 1998; Peebles \& Ratra 1988) are the
most discussed candidates.  The effective equation of state of the
dark energy is usually parameterized as $w=p_Q/\rho_Q$, where $w=-1$
for a pure cosmological constant but may be larger for a scalar field
component or smaller for phantom energy (Caldwell \etal 2003); $w$ may be
constant or time-varying.  If $w$ is assumed to be constant in time,
the combined data sets from Type Ia supernovae (SNe) and the CMB or
large-scale structure already imply $-1.30 \le w \le -0.6$ (Spergel
\etal 2003; Baccigalupi \etal 2002; Wang \etal 2000).  Our calculations
apply to models with a constant or slowly varying equation of state
(Steinhardt, Wang, \& Zlatev 1999).  We generically refer to dark
energy as a ``Q-component'', and to cosmological models with a mixture
of dark energy and cold dark matter (CDM) as QCDM.  \lcdm is then a
special case of QCDM with $w=-1$.

\subsection{Cosmography} 
\label{sec:cosmography}

In the limit of a nearly homogeneous Q-component, all the consequences 
of the Q-field follow from its effect on the expansion rate:
\begin{equation}
  \frac{H^2(z)}{H_0^2} = \Omega_M(1+z)^3+\Omega_Q(1+z)^{3(1+w)}+\Omega_K(1+z)^2,
\end{equation}
where $H_0$ is the Hubble parameter today, $\Omega_M$ and $\Omega_Q$
are the matter and dark energy density parameters today, and
$\Omega_K=1-\Omega_M-\Omega_Q$ is the contribution of curvature. 
In an $\Omega_K=0$ flat universe, the angular diameter distance from
an object at redshift $z_1$ to an object at redshift $z_2$ is given
by 
\begin{equation}
  D_A(z_1,z_2) = \frac{c}{1+z_2} \int_{z_1}^{z_2} \frac{dz}{H(z)}\ .
\end{equation}
 
\subsection{Linear growth of perturbations}

In first-order perturbation theory, the growing-mode solution of the
matter density contrast satisfies the differential equation

\be \ddot
D + 2H(z)\dot D = \frac{3}{2} H_0^2 \Om_M (1+z)^3 D \,, \label{eq:D} 
\ee

where $D$ is the linear growth factor and the dots denote derivatives
with respect to time.  An analytical solution in terms of
hypergeometric functions can be found in Padmanabhan (2003). A good
approximation to the instantaneous growth index $f = d\ln D/d\ln a$ as
the matter density parameter $\Omega_M(z)=\Omega_M(1+z)^3(H_0^2/H^2)$
approaches unity is given by

\be f\approx \Omega_M(z)^{{3-3w\over5-6w}}=\left[{\rho_M(z)\over \rho_M(z)+\rho_Q(z)}\right]^{{3-3w\over 5-6w}},
\ee

where $\rho_Q$ and $\rho_M$ are the Q-field and matter densities,
respectively (Silveira \& Waga 1994; Wang \& Steinhardt 1998).  We see
that the growth of density perturbations is slowed as $\rho_Q$
approaches $\rho_M$.  Growth suppression increases with increasing $w$
since the onset of dark energy domination occurs at higher redshift:
in order to achieve the level of mass fluctuations observed today,
cosmic structure in QCDM must form at earlier times (Huterer \& Turner
2001).
 
\subsection{Power spectrum}

Dark halos form from primordial matter density fluctuations in the
early universe.  The Q-field is so light that on scales less than
a few hundred Mpc fluctuations in Q disperse relativistically, and
quintessence behaves as a smooth component like the cosmological
constant.  On very large scales, however, the dark energy may
cluster gravitationally.  Fluctuations in the Q-field add to the
right-hand side of of equation (\ref{eq:D}), resulting in a
different growth rate once quintessence starts to dominate the
cosmological energy density.  This change of behavior is
incorporated in the transfer function for the matter density
field, which accounts for all modifications of the primordial
power-law spectrum due to the effects of pressure and dissipative
processes, together with the gravitational clustering of matter
and dark energy.  The linear power spectrum for matter density
perturbations in QCDM scenarios can be expressed as
\be \label{eq:Pk}
  P(k,z;w) = A k^n T_Q^2(k,z;w) \left[\f{D(z;w)}{D_0}\right]^2 ,
\ee
where $A$ is an overall normalization, $k$ is the wavenumber, $n$
is the spectral index of the post-inflationary adiabatic density
perturbation, $T_Q(k,z;w)$ is the transfer function, and $D$ is the
growth factor from equation (\ref{eq:D}). 

Ma \etal (1999) provide a prescription in which the modification due
to quintessence with $w>-1$ are captured by a multiplicative factor
$T_{\rm Q\Lambda}(k,z)$ to be applied to the standard \lcdm transfer
function $T_\Lambda(k)$, i.e.\ $T_Q \equiv T_{\rm Q\Lambda} T_\Lambda$.
(The \lcdm transfer function can be taken from, e.g., Eisenstein \& Hu
1999.)  In contrast to models with no dark energy ($\Om_Q=0$) or ones
with a nonzero cosmological constant, ($\Om_Q \ne 0$, $w=-1$), a
Q-field with $\Om_Q>0$, $w>-1$ adds power to $P(k)$ at small
wavenumbers, increasingly so as the contribution of quintessence to
the energy density becomes larger.  This effect introduces an additional
dependence of the transfer function on redshift (Ma \etal 1999).  At
this time no such prescription for $w<-1$ phantom energy type models
has been published.  Fortunately the modifications to the \lcdm
transfer function are non-negligible only on scales comparable to
the horizon.  On the sub-cluster scales relevant for our calculations
it is thus possible to ignore these effects.

\subsection{Mass variance}
\label{sec:sigmaM}

On the scale of galaxy clusters and below, the QCDM power spectrum
has identical shape to the $\Lambda$CDM one and differs only in the
overall amplitude.  To fix the normalization factor $A$ in equation
(\ref{eq:Pk}), we compute the variance of the mass-density field
smoothed on a scale of comoving radius $R$,  
\be \label{eq:var}
  \s_M^2 = \langle(\delta M/M)^2\rangle
  = \frac{1}{2 \pi^2} \int_0^\infty dk\, k^2 P(k,z) W^2(k R) \,,
\ee
where $M = H_0^2 \Om_M R^3/2G$ is the mass inside a sphere of
radius $R$, and $W(k R)$ is the Fourier transform of the spherical
top-hat window function, $W(x)=(3/x^2)(\sin x/x-\cos x)$.  For
calculations of lensing by halos on the scale of clusters and
smaller, it seems most appopriate to normalize the QCDM power
spectrum by the local abundance of galaxy clusters.\footnote{By
contrast, Sarbu \etal (2001) compute lens statistics using a QCDM
power spectrum normalized by the COBE measurement of the cosmic
microwave background, corresponding to scales near the horizon
scale today.} Although the normalization of a QCDM power spectrum
by cluster abundances has been discussed by Wang \& Steinhardt
(1998), we present it in the Appendix since we use the more recent
data set of Reiprich \& B\"ohringer (2002).  We find that the data
favor a value of the rms mass fluctuations 8~$h^{-1}$~Mpc spheres,
$\s_8 \equiv \sigma(z=0,R=8\,h^{-1}\,{\rm Mpc}) = 0.74$, for
the currently favored cosmology with $\Omega_M=0.3$, $\Om_Q=0.7$,
$h=0.7$, and $n=1$, independently of $w$ for $-1.5<w<-0.5$ (see
the Appendix for details).  Then the normalization factor in
eq.~(\ref{eq:Pk}) is
\be
A = \frac{2\pi^2\s_8^2}{\int_0^\infty dk\,k^{n+2} T_Q^2(k,0) W^2 (k R_8)}\ .
\ee 

Following Bullock \etal (2001) we define $M_*$ as the mass of halos
collapsing today from 1-$\sigma$ fluctuations, \ie as the solution of
$\s_M=\de_c$, where $\de_c$ is the linearly extrapolated overdensity
at collapse (see \S \ref{sec:tophat}). Table~\ref{tab:mstar}
shows $M_*$ for various values of $w$, for our favored value of
$\s_8=0.74$ and also for $\s_8=0.90$.

\begin{deluxetable}{cccccc}
\tablecaption{$M_*(w) \; [10^{12}\msun]$}
\tablewidth{0pt}
\tablehead{
 $\s_8$ &         &         &   $w$   &         &         \\
        & $-0.50$ & $-0.75$ & $-1.00$ & $-1.25$ & $-1.50$
}
\startdata
 $0.74$ & 3.71 & 3.58 & 3.52 & 3.47 & 3.44 \\
 $0.90$ & 11.9 & 11.5 & 11.3 & 11.2 & 11.1 \\
\enddata
\label{tab:mstar}
\end{deluxetable}
 
\subsection{Halo mass function} 
\label{sec:massf}

The mass function of collapsed objects has been derived from
theoretical arguments and from $N$-body simulations.  The classic
prediction comes from Press \& Schechter (1974), and it has been
adjusted by Sheth \& Tormen (1999) to better fit results from
simulations.  Jenkins \etal (2001) have presented an alternate
fitting form that they claim fits a wide variety of $N$-body
simulations.  The functional forms of these three models are:
\begin{subequations}
\begin{eqnarray}
n_{\rm PS}(M)&=&-\sqrt{\f{2}{\pi}} \f{\rho_b}{M} \f{\de_c}{\s_M^2} \f{d\s_M}{dM} \exp\left[-\f{1}{2}\left(\f{\de_c}{\s_M}\right)^2\right] \\
n_{\rm ST}(M)&=&-\f{0.383}{\sqrt{\pi}} \f{\rho_b}{M} \f{\de_c}{\s_M^2} \f{d\s_M}{dM} \left[1+\left(\f{\s_M^2}{0.707 \de_c^2} \right)^{0.3} \right] \nonumber \\
&&\times \exp\left[-\f{0.707}{2} \left(\f{\de_c}{\s_M}\right)^2\right] \\
n_{\rm J}(M)&=&-\f{\rho_0}{M} \f{d\s_M}{dM} \f{0.315}{\s_M} \exp\left[-|0.61-\ln \s_M|^{3.8}\right].
\label{eq:massfunctions}
\end{eqnarray}
\end{subequations}
We adopt the Jenkins \etal mass function as our fiducial model, but
we explore the effects of using the other approximations as well.
While the original analysis by Jenkins \etal did not include any
cases with $w \ne -1$, recent results from $N$-body simulations
including the effect of a dark energy component with $w \ne -1$
confirm the validity of the fitting function (Linder \& Jenkins
2003; Kuhlen \etal, in preparation).

\subsection{Spherical top-hat collapse} 
\label{sec:tophat}

The PS and ST mass functions and the definition of $M_*$ depend on the
linearly extrapolated overdensity at collapse $\de_c$.  The SIS and
NFW halo lensing cross sections depend on the virial overdensity
$\De_{\rm vir}=(\rho/\rho_M)_{\rm vir}$.  Both $\de_c$ and $\De_{\rm
vir}$ are determined by solving the spherical collapse model (e.g.,
Wang \& Steinhardt 1998; Strigari 2003, private communication).  While
the changes in $\de_c$ with $w$ are only a few percent, the changes in
$\De_{\rm vir}$ are more noticeable.  As $w$ increases $\De_{\rm vir}$
becomes larger as structures form earlier.  At $z=0$ we find $\De_{\rm
vir}=(499, 409, 346, 298, 270)$ for $w=(-0.50, -0.75, -1.00, -1.25,
-1.50)$ (Strigari 2003, private communication).  For $w \ge -1$
fitting formulae for both quantities have been published by Weinberg
\& Kamionkowski (2002).

\section{Understanding the Parameters}
\label{sec:ParamDep}

\begin{figure}
\plotone{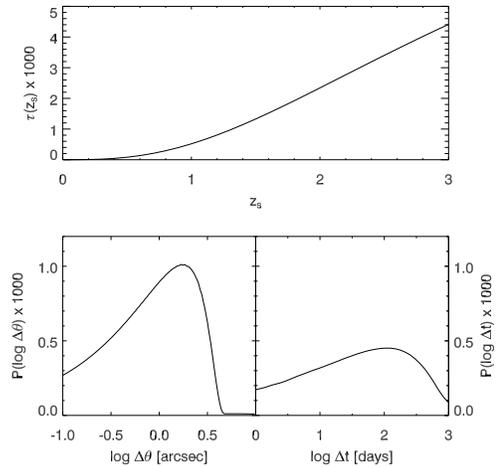}
\caption{
Lens statistics results for the fiducial model.
(Top) The total lensing probability as a function of the source
redshift.
(Bottom) The image separation and time delay distributions.
}
\label{fig:fiducial}
\end{figure}

\begin{figure*}
\includegraphics[width=7in]{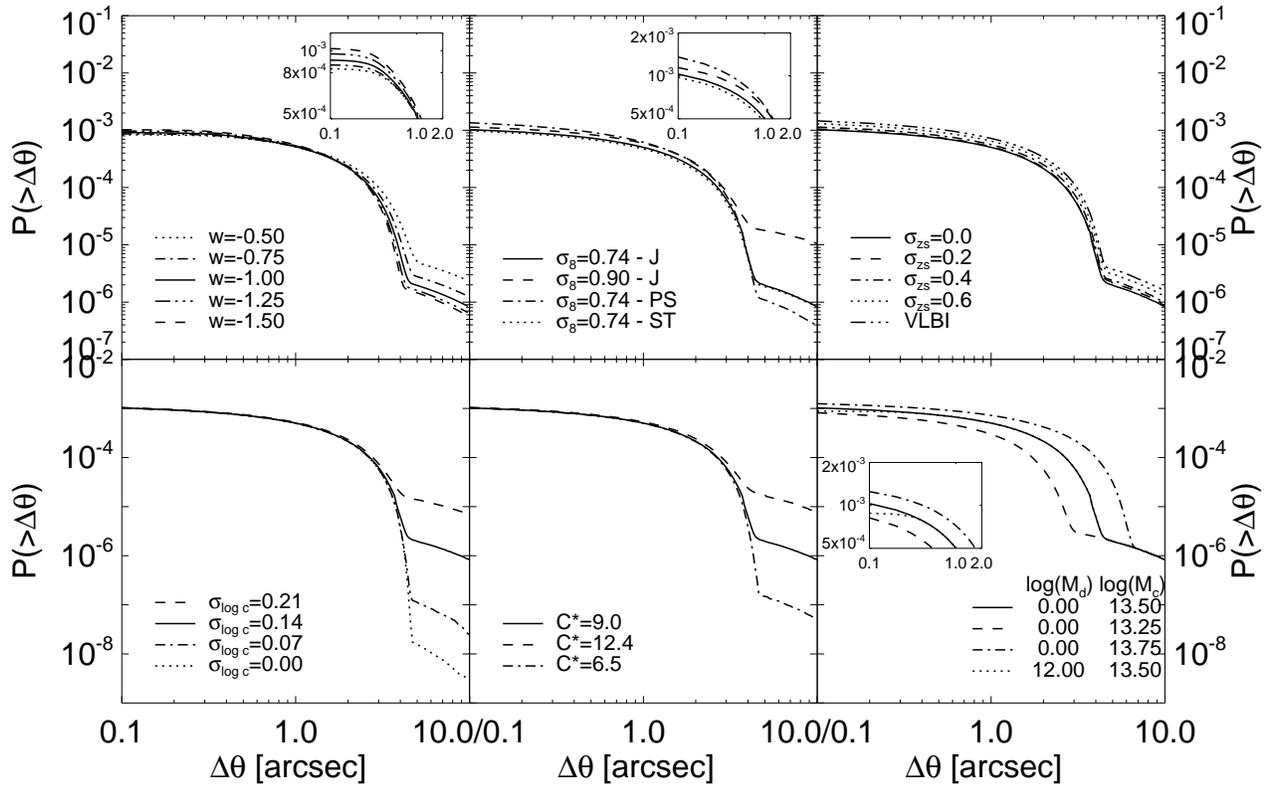}
\caption{The effects of parameter variations on the cumulative
distribution of image separations.
}
\label{fig:sep}
\end{figure*}

\begin{figure*}
\includegraphics[width=7in]{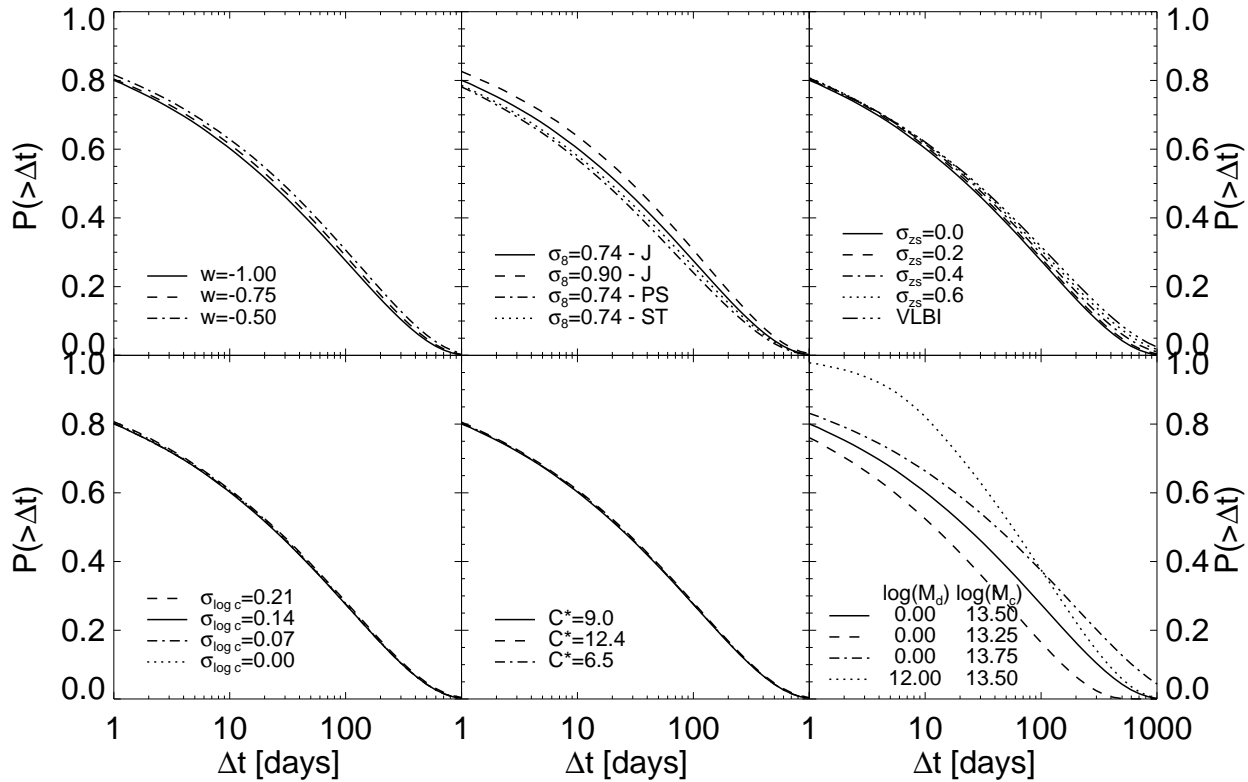}
\caption{The effects of parameter variations on the normalized
cumulative distribution of time delays.}
\label{fig:dt}
\end{figure*}

\begin{figure*}
\includegraphics[width=7in]{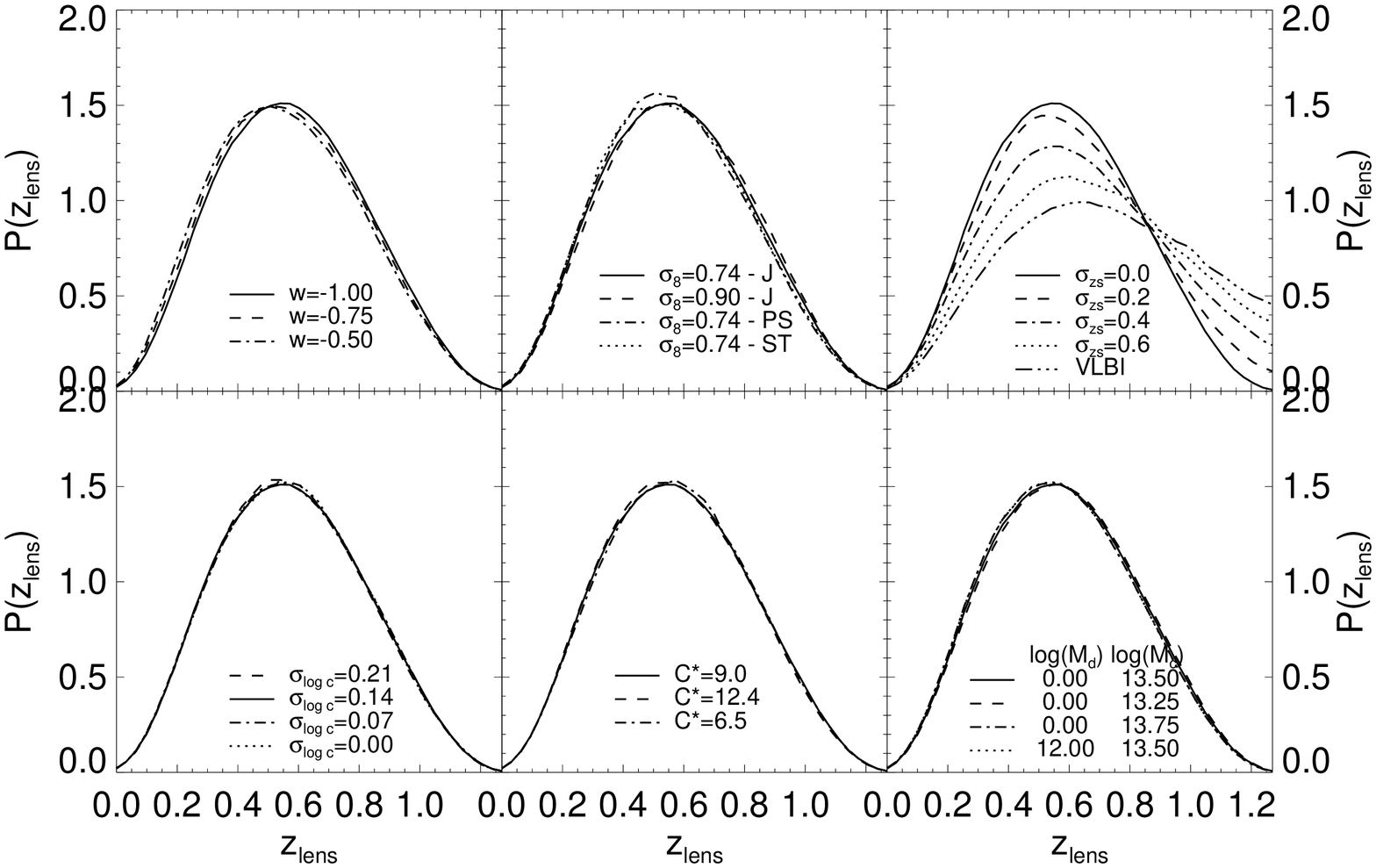}
\caption{The effects of parameter variations on the normalized
differential distribution of lens redshifts.}
\label{fig:zl}
\end{figure*}

From the previous two sections it is clear that there are a number
of parameters that enter into lens statistics calculations.  In this
section we study how they affect the results.  Our goal is not to
exhaustively catalog and quantify all parameter dependences and
covariances, but to identify the main qualitative features.  To do
this, we adopt a fiducial model (which shall be justified in
\S\ref{sec:CLASScomp}), and then vary the parameters one at a time.
The fiducial model and the variations are summarized in
Table~\ref{tab:fiducial}.  Figure~\ref{fig:fiducial} shows the
optical depth as a function of source redshift, the image separation
distribution, and the time delay distribution for the fiducial model.
Figures~\ref{fig:sep}--\ref{fig:zl} then show how the parameter
variations affect the image separation distribution, the time delay
distribution, and the lens redshift distribution.

Note that the image separation distributions are normalized so the
total probability reflects the probability of lensing, while the
time delay and lens redshift distributions are normalized so the
total probability is unity.  This allows us to separate changes in
the overall lensing probability from changes in the shapes of the
time delay and lens redshift distributions.  Also note that in
Figures~\ref{fig:sep} and \ref{fig:dt} we plot the cumulative
separation and time delay distributions, but for completeness in
Figure~\ref{fig:fiducial} we show the differential distributions.
Finally, note that the calculations in this section are designed
for comparison with the CLASS survey: we use a source redshift
distribution with mean $\langle z_s \rangle = 1.27$, and we compute
magnification bias using the total magnification and a source
luminosity function modeled as a power law $dN/dS \propto S^{-2.1}$
(see Rusin \& Tegmark 2001; Browne \etal 2002).

\begin{deluxetable}{ccc}
\tablecaption{Fiducial Model and Variations}
\tablewidth{0pt}
\tablehead{
 Parameter & Fiducial Value & Variation Values
}
\startdata
$\Om_m$     & $0.3$   & -- \\
$\Om_q$     & $0.7$   & -- \\
$w$         & $-1.00$ & $-0.50, -0.75, -1.25, -1.50$ \\
$n$         & $1.0$   & -- \\
$h$         & $0.7$   & -- \\
$dn/dM$     & Jenkins & PS, ST \\
$\medc$     & $9.0$   & $6.5, 12.4$ \\
$\sigc$     & $0.14$  & $0.21, 0.07, 0.00$ \\
$\log\Md$   & $0.0$   & $12.0$ \\
$\log\Mc$   & $13.50$ & $13.25, 13.75$ \\
$\s_{z_s}$  & $0.0$   & $0.2, 0.4, 0.6, {\rm VLBI}$ \\
\enddata
\label{tab:fiducial}
\end{deluxetable}

\subsection{The fiducial model}

In our fiducial model, the image separation distribution peaks
around $\Dth \sim 1\arcsec$--$2\arcsec$, and has a long tail to
small separations (see Fig.~\ref{fig:fiducial}).  Above the peak
the distribution drops quickly, then has a kink at
$\Dth \sim 4\arcsec$ with a long but low-amplitude tail to large
separations.  This feature represents the transition from normal
lenses produced by galaxies modeled as SIS halos, to wide-separation
lenses produced by groups and clusters modeled as NFW halos.  NFW
halos are much less efficient lenses than SIS, so wide-separation
lenses are much less abundant than normal arcsecond-scale lenses.

The time delay distribution is surprisingly broad, spanning more
than three orders of magnitude (also see Oguri \etal 2002).  Thus,
in future surveys a substantial fraction of lenses will have time
delays on week or month time scales for which measurements will be
feasible, but a non-negligible fraction of lenses will have time
delays that are too long to be of practical use.  There is actually
a kink in the time delay distribution marking the transition from
normal to wide-separation lenses, but it occurs at $\sim$1100~days
and is not visible in Figure~\ref{fig:fiducial}.

\subsection{Variation in $w$}

As discussed in \S\ref{sec:StrForm}, the fact that dark energy
modifies cosmological dynamics is taken into account self-consistently
in everything except for the effect on the NFW concentration
parameter.  Since overdensities collapse earlier as $w$ increases, the
median concentration should increase with $w$.  Some analytical and
numerical studies of halo concentrations in universes with $w \ne -1$
have been made (Bartelmann, Perrotta, \& Baccigalupi 2003; Klypin
\etal 2003; Kuhlen \etal, in preparation), but so far no $\medc(w)$
relations have been published.  The explicit $w$ dependence of
$\medc(w)$ appears to be rather small, certainly less than the scatter
$\sigc$, so we neglect it in this work.

Figure~\ref{fig:sep} shows that changing $w$ affects the full
range of image separations.  The effects are largest at wide
separations ($\Dth \gtrsim 4\arcsec$), with more than twice as many
lenses for $w=-0.50$ as for \lcdm.  This makes sense, because
wide-separation lenses are produced by massive group and cluster
halos, which are more abundant at redshifts $z_l \sim 0.2$--$1$
in models with larger $w$ by virtue of the earlier structure
formation.  Increasing $w$ slightly decreases the number of
lenses with $\Dth \lesssim 2\arcsec$; among the parameters we
examine, this is the main change at arcsecond scales.  An
increase in $w$ also shifts the time delay distribution to
slightly higher values (Fig.~\ref{fig:dt}), and shifts the lens
redshift distribution to slightly lower values (Fig.~\ref{fig:zl}).

We conclude that wide-separation lenses can be used to place
useful constraints on $w$ only after parameters governing the
abundance and structure of massive halos ($\s_8, \medc, \sigc,
\Mc$) are much better constrained (cf.\ Sarbu \etal 2001).  Normal
arcsecond-scale lenses can provide more robust constraints on $w$,
but large future samples will be needed to detect the relatively
weak effects of $w$.  In this regard our results differ from
those of Huterer \& Ma (2003), who claim an upper limit $w<-0.85$
at 68\% confidence using the current CLASS lens sample.

\subsection{Variation in the mass function and $\s_8$}

The strongest effect in this group comes from changes in $\s_8$.
In order to match the observed abundance of clusters from HIFLUGCS
we have adopted a low value $\s_8 = 0.74$.  Determinations of $\s_8$
at much larger scales (COBE, WMAP, weak lensing, excess power in CMB
at $l > 1000$, etc.; Ma \etal 1999; Spergel \etal 2003; Huterer \&
White 2002; Komatsu \& Seljak 2002), by contrast, tend to favor a
higher value $\s_8 \sim 0.90$.  At present there exists no satisfying
reconciliation of this tension, so we are forced to treat $\s_8$ as
an unknown.  The seemingly small increase in $\s_8$ from $0.74$ to
$0.90$ produces a change of about an order of magnitude in the
abundance of wide-separation lenses.  The total lensing probability,
however, is affected by only 10\%.  The increased abundance of
wide-separation lenses, which have long time delays, shifts the time
delay distribution to slightly higher values.

The use of different mass functions, Press--Schechter vs.\
Sheth--Tormen vs.\ Jenkins, produces a smaller effect.  The PS mass
function tends to over- (under-) predict the abundance of halos
below (above) $\sim\!10^{14} \msun$ relative to $N$-body simulations,
while the ST and Jenkins mass functions both aim to correct this
problem.  Thus it is not surprising to see that the PS mass function
tends to predict more small-separation lenses and fewer large-separation
lenses than the ST/J mass functions.  Relative to the Jenkins mass
function, the ST mass function produces slightly fewer lenses at all
separations, but the differences are less than 10\%.  The tendency
for the PS and ST to shift the image separation distribution to lower
values also shifts the time delay distribution to lower values; the
shift in the median is 17\% for ST and 25\% for PS.

\subsection{Variation in source distribution}

Our fiducial model assumes that all sources lie at redshift $z_s=1.27$,
but we consider several other possibilities as follows.  First, we
consider Gaussian distributions with the same mean but different widths
$\sigma_{z_s}$.  Second, we consider using the source redshift distribution
from the 2nd Caltech/Jodrell Bank VLBI sample (Henstock \etal 1997).  This
distribution is reasonably well described by a Gaussian with mean
$\langle z_s \rangle = 1.33$ and width $\sigma_{z_s} = 0.75$, but we use
the actual observed distribution.  This is the same redshift distribution
used by Sarbu, Rusin, \& Ma (2001) in their study of the effects of dark
energy on strong lensing.

Figure 2 shows that with the mean source redshift fixed, increasing the
scatter even to $\sigma_{z_s} \sim 0.4$ has little effect on the optical
depth or image separation distribution.  Only when the scatter reaches
$\sigma_{z_s} \sim 0.6$ is there a notable increase, which is due to the
growing number of sources at higher redshifts (where the optical depth is
higher).  The higher optical depth for the model with the VLBI redshift
distribution ($\sim$40\% higher than for the fiducial model) is due partly
to the higher mean redshift and partly to the large width of the
distribution.  Not surprisingly, broadening the source redshift
distribution causes the lens redshift distribution to show a larger tail
to higher redshifts.

\subsection{Variation in $\medc$ and $\sigc$}

The parameters $\medc$ and $\sigc$ control the distribution of
concentration parameters for NFW halos.  They have little effect on
arcsecond-scale lenses because those are produced by galaxies with
roughly isothermal profiles.  But they have an enormous effect on
the abundance of wide-separations lenses produced by group and
cluster halos.  Increasing the median concentration $\medc$ by
0.14~dex (the 1$\sigma$ scatter found by Bullock \etal 2001)
increases the abundance of wide-separation lenses by more than a
factor of 8, while decreasing $\medc$ by 0.14~dex decreases the
abundance by a factor of $\sim$15.  Even more extreme is the effect of
changing the scatter in the $c(M,z)$ relation.  Neglecting the
scatter (using $\sigc=0$) reduces the abundance of wide-separation
lenses by more than two(!) orders of magnitude. Both effects are explained
by the fact that making a halo less concentrated can dramatically
reduce its lensing cross section and its ability to produce lenses
of a given size.  This, of course, was the original reason for
using NFW halos in lens statistics: halos that are less concentrated
than SIS are needed to explain the lack of observed wide-separation
lenses.

\begin{figure}
\plotone{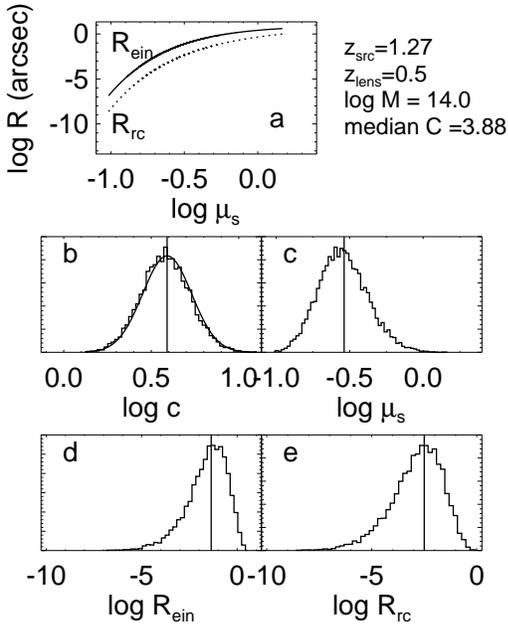}
\caption{
The change in the lensing properties of an NFW halo with concentration.
(a) The strong dependence of the Einstein radius $R_{\rm ein}$ and
the position of the radial caustic $R_{\rm rc}$ on the dimensionless
lensing efficiency parameter $\mu_s$.
(b) A log-normal distribution of concentrations with width
$\sigc=0.14$~dex.  The histogram shows the distribution of values
drawn from the parent distribution represented by the smooth curve.
(c--e) The resulting distributions of the efficiency parameter $\mu_s$,
the Einstein radius $R_{\rm ein}$, and the radial caustic $R_{\rm cr}$.
The vertical lines indicate the values corresponding to the median
concentration.
}
\label{fig:NFW_c_and_sigma}
\end{figure}

It is important to understand this result in more detail. In
Figure~\ref{fig:NFW_c_and_sigma} we show how varying the concentration
(for fixed mass and redshift) affects the lensing properties of an NFW
halo.  Panel (b) shows a log-normal distribution of concentrations
with width $\sigc=0.14$~dex.  Panel (c) shows the corresponding
distribution of the lensing efficiency parameter $\mu_s$. When the
mass is fixed, varying the concentration changes $r_s$ and $\rho_s$,
and hence changes the lensing strength parameter $\mu_s$.  Because of
the steep dependence on $\mu_s$ (panel a), small changes in the
concentration can produce large changes in the Einstein radius and
radial caustic.  As a result, a 0.14 dex scatter in concentrations
leads to a spread of some five orders of magnitude in the scale radii
(panels d and e), or ten orders of magnitude in the lensing cross
section ($\sigma^*_{\rm NFW} \propto R_{\rm rc}^2$).

\begin{figure}
\plotone{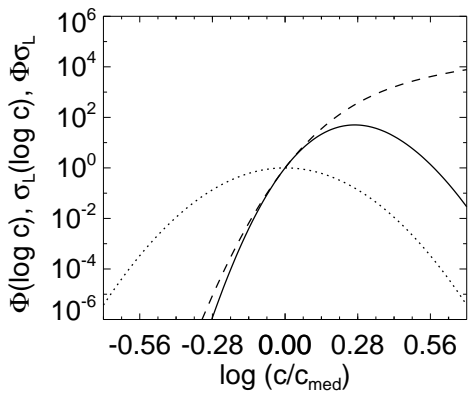}
\caption{ $\sigma_L$, the lensing cross section (dashed), $\Phi$, a
Gaussian distribution in $\lg(c)$ with width $\sigc=0.14$~dex (dotted), and the product of the two
(solid), for the same lens-source system as in
Figure~\ref{fig:NFW_c_and_sigma}. All curves have been normalized to
unity at $c=c_{\rm med}$. Since $\sigma_L$ is a steeply rising
function, the product of $\sigma_L\Phi$ peaks at around two standard
deviations about the median, with a weighted cross section about two
orders of magnitude larger than at $c_{med}$.
}
\label{fig:sigmac_gauss}
\end{figure}

In Figure~\ref{fig:sigmac_gauss} we have plotted the lensing cross
section $\sigma_L(\lg c)$, the Gaussian distribution of the
concentration $\Phi(\lg c)$, and their product. All curves are
normalized to unity at $c=c_{\rm med}$. The increase in the lensing
cross section with concentration outweighs the decrease in the
Gaussian tail out to 4 standard deviations. A peak occurs at around
$\lg(c/c_{\rm med})/\sigc=2$, where the product of $\sigma_L$ and
$\Phi$ is about two orders of magnitude larger than at $c_{\rm
med}$. The weighted cross section is given by the integral over this
product ($\int \sigma_L \Phi \, d\!\log c / \int \Phi \, d\!\log c$),
which explains the difference between a single NFW halo at $c=c_{\rm
med}$ and a distribution of NFW halos around this concentration with
scatter equal to $\sigc$.

Put another way, even a massive halo with the median concentration
has a relatively small Einstein radius and cross section, so if all
halos have the median concentration ($\sigc=0$) then wide-separation
lenses will be quite rare.  Only the most concentrated halos have
reasonable cross sections for producing wide-separation lenses.
Thus, wide-separation lenses are expected to be abundant only if
the median and/or scatter in the concentration distribution is
large enough to produce a sizable population of massive, concentrated
halos.  We conclude that it is imperative to include the distribution
of concentrations in lens statistics calculations --- and that the
main limitation in predictions of wide-separation lenses will be
uncertainties in $\medc$ and $\sigc$.  These effects have been
mentioned before (e.g., Keeton \& Madau 2001; Wyithe, Turner, \&
Spergel 2001), but we were not previously aware that they could
affect NFW lensing so dramatically.

\subsection{Variation in $\Md$ and $\Mc$}
\label{sec:MdMc_var}

The upper transition mass $\Mc$ parametrizes the change from
galaxy-sized SIS to group- and cluster-sized NFW lenses, and its
effect on the image separation distribution is orthogonal to the
effects of the other parameters.  Specifically, whereas the main
effects discussed so far are increases or decreases in the abundance
of wide-separation lenses, what $\Mc$ does is determine the {\it
position\/} of the kink in the image separation distribution that
marks the transition from normal to wide-separation lenses.  For
our fiducial value $\log\Mc=13.50$ the transition occurs at
$\Dth \approx 4\arcsec$, while for $\log\Mc=13.25$ it shifts down
to $\Dth \approx 3\arcsec$ and for $\log\Mc=13.75$ it shifts up
to $\Dth \approx 6\arcsec$.  The lack of lenses with separations
larger than $3\arcsec$ in the CLASS statistical sample makes it
possible to place upper limits on $\Mc$ (see \S\ref{sec:CLASScomp}).

The lower transition mass $\Md$ parametrizes the change from
dwarf-sized NFW to galaxy-sized SIS lenses.  Making $\Md$ non-zero
causes a drop in the abundance of lenses (in direct analogy with
the galaxy to cluster transition at $\Mc$).  But the effects are
not readily apparent in the image separation distribution, because
they occur only at the smallest image separations.  Except for very
large values $\log\Md \sim 12$, the effect is confined to
$\Dth \lesssim 0\farcs1$.  Because today's largest statistically
complete lens sample is complete only above $0\farcs3$, the lower
transition mass $\Md$ is not well constrained by the available
data (see \S\ref{sec:CLASScomp} for details).  For this reason
we have chosen to omit the lower transition from our fiducial
model.

Even more interesting are the effects of the transition masses on
the time delay distribution.  Increasing $\Mc$ effectively
increases the abundance of lenses with intermediate separations
$\Dth \sim 4\arcsec$--$6\arcsec$, which have long time delays; so
it shifts the time delay distribution to higher values.  Applying
the $\Md$ transition reduces the abundance of lenses with small
separations, so it removes many of the lenses with short time
delays and hence changes the shape of the time delay distribution.
Of all the parameter variations we consider, $\Mc$ and $\Md$ have
the most significant effects on the time delay distribution.

\subsection{General comments}

In previous analyses it was not clear whether lens statistics
could simultaneously constrain both the cosmology and the profiles
of massive dark matter halos.  Our results suggest that the answer
is yes, because of information contained in the distribution of
lensed image separations (also see Huterer \& Ma 2003).  The
abundance of wide-separation lenses depends primarily on $\s_8$
and the core properities of NFW halos (the distribution of
concentrations; or equivalently, in generalized NFW models, the
distribution of inner cusp slopes).  The exact location of the break
at $\Dth \sim 4\arcsec$ between common arcsecond-scale lenses
and less abundant wide-separation lenses probes baryonic cooling
processes that determine why galaxies and clusters have different
density profiles.  The time delay distribution depends on these
processes as well.  Finally, the effects of $w$ are seen at all
image separations.  Future data should therefore be able to break
degeneracies between parameters by considering not only the total
number of lenses but also the separation distribution.

\section{Comparisons with the JVAS/CLASS Survey}
\label{sec:CLASScomp}

Already we can place constraints on some of the parameters using
existing lens samples.  The largest homogeneous sample comes from
the combination of the Jodrell Bank/VLA Astrometric Survey (JVAS)
and Cosmic Lens All-Sky Survey (CLASS), which in its entirety
comprises VLA observations of 16503 radio sources with 22
multiply-imaged systems (Browne \etal 2002).  A subset containing
8958 sources and 13 lenses forms a statistically complete sample
with well-defined selection criteria (Myers \etal 2002).  The
survey is believed to be complete at $\Dth \ge 0\farcs3$; the
observed image separations range from $0\farcs33$ to $4\farcs55$
in the full sample, and $0\farcs33$ to $2\farcs57$ in the
statistical subsample.  The source redshift distribution is not
well known, but the mean redshift of a subsample of sources is
$\langle z_s \rangle = 1.27$ (Marlow \etal 2000).

In the range of angular separations probed by JVAS/CLASS the
transition masses $\Md$ and $\Mc$ are the most sensitive parameters.
Variations in $\medc$ and $\sigc$ have larger effects overall, but
only in the wide-separation tail where the null results of the
JVAS/CLASS survey precludes interesting constraints.  While Huterer
\& Ma (2003) claim to be able to distinguish different values of
$w$ and $\s_8$ using the CLASS statistical sample, we find
that the effects of those parameters are smaller than the Poisson
uncertainties.  We therefore restrict attention to $\Md$ and $\Mc$.

For a statistical analysis of the {\it shape\/} of the image
separation distribution, it seems reasonable to use all 22
lenses.  The incompleteness of the full sample should not affect
the normalized separation distribution.  We quantify the agreement
between our models and the data using the Kolmogorov--Smirnov (K.S.)
test (e.g., Press \etal 1992, \S14.3).  The K.S.\ statistic is
\be
  D_{\rm KS} = \mathop{\rm max}_{0<\Dth<\infty}
    |P_{\rm obs}(<\!\Dth) - P_{\rm mod}(<\!\Dth)| ,
\ee
where $P_{\rm obs}(<\!\Dth)$ and $P_{\rm mod}(<\!\Dth)$ are the
observed and predicted cumulative image separation distributions.
Associated with $D_{\rm KS}$ is the K.S.\ probability $P_{\rm KS}(D)$,
quantifying the significance of an observed value of $D$ as a disproof
of the hypothesis that the data and the model represent the same
distribution.  Large values of $D_{\rm KS}$ correspond to small
values of $P_{\rm KS}$ and indicate poor agreement between data and
model.

\begin{figure}
\plotone{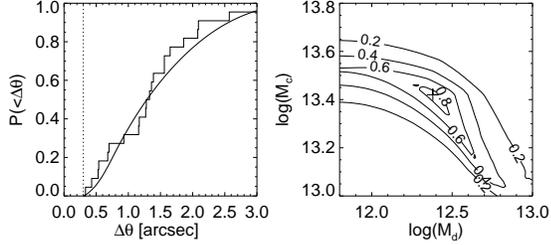}
\caption{Comparison with JVAS/CLASS by K.S.\ test. The histogram on
the left shows the normalized cumulative image separation distribution
for all 22 lenses in the JVAS/CLASS sample.  The solid line shows our
best-fit model, determined by choosing $\Md$ and $\Mc$ to minimize
the K.S.\ measure $D_{\rm max}$.  The dotted vertical line indicates
the JVAS/CLASS completeness limit, $\Dth=0\farcs3$. The right panel
shows contours of $P_{\rm KS}$ with the maximum at $P_{\rm KS}=0.887$ marked by the cross.}
\label{fig:kstest}
\end{figure}

Figure~\ref{fig:kstest} shows contours of $D_{\rm KS}$ versus the
transition masses $\Md$ and $\Mc$.  The best fit (minimum value
of $D_{\rm KS}$) occurs at $\log\Md=12.38$ and $\log\Mc=13.42$,
and has a probability $P_{\rm KS}=88.7\%$.  This value of the
galaxy/cluster transition mass agrees well with the value found by
Huterer \& Ma (2003) from a similar analysis of the CLASS sample.
The large inferred value of $\Md$ is surprising; if taken at face
value, it implies that even a relatively massive galaxy like the
Milky Way is a ``dwarf'' whose density profile is NFW.  It is not
clear whether this result reflects a shortcoming of our theoretical
modeling, or an unidentified selection effect in the JVAS/CLASS
sample, or some other effect.

The K.S.\ test applies only to the shape of the image separation
distribution and is insensitive to the total lensing probability.
As a complementary approach, we use a $\chi^2$ test to compare the
unnormalized differential distributions; this test is somewhat less
sensitive to the shapes of the distributions, but very sensitive to
the relative normalizations.  We take care to use only the statistically
complete JVAS/CLASS subsample.  We bin the 13 lenses into seven bins
of width $\log\Dth=0.2$; although the two right-most bins are empty,
we include them because our models can predict non-negligible numbers
of lenses in this regime.

\begin{figure}
\plotone{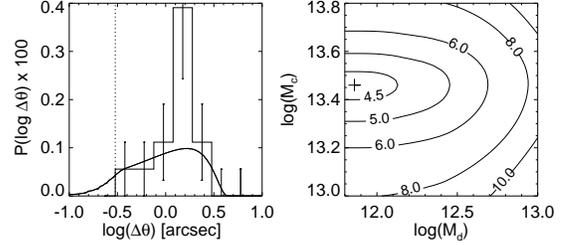}
\caption{Comparison with JVAS/CLASS data by $\chi^2$ test.  The
histogram on the left shows the unnormalized differential image
separation distribution for the 13 lenses in the statistically
complete JVAS/CLASS subsample.  The error bars represent Poisson
noise.  The solid line shows our best-fit model, determined by
choosing $\Md$ and $\Mc$ to minimize $\chi^2$.  The dotted vertical
line indicates the JVAS/CLASS completeness limit for $\De \th$.
The right panel shows contours of $\chi^2$ with the minimum marked
by the cross.}
\label{fig:chi2}
\end{figure}

Figure~\ref{fig:chi2} shows contours of $\chi^2$ versus the transition
masses.  The best-fit model has $\chi^2=4.39$ for seven bins.  We
obtain good constraints on the upper transition mass $\Mc$, with a
best-fit value of $\log\Mc=13.46$ in good agreement with the results
of the K.S.\ test.  For $\Md$, by contrast, the $\chi^2$ surface
flattens into a long valley at $\log\Md<12$. We are therefore unable
to place a strong constraint on $\Md$.  At $68\%$ confidence level
we can only place an upper bound of $\log\Md<12.57$.  As mentioned in
\S\ref{sec:MdMc_var}, the effect of lowering $\Md$ below $10^{12}\msun$
is confined to very small angular separations $\log\Dth \lesssim -0.5$,
outside of the range probed by JVAS/CLASS.  Future surveys extending
to smaller image separations will be needed to robustly constrain
$\Md$.

We note that although our models produce an acceptable value of
$\chi^2$, they cannot account for the sharp peak in the observed
distribution at $\log\Dth \sim 0\farcs2$.  The presence of this peak
actually explains why the K.S.\ test produces such a large best-fit
value for $\Md$: being sensitive only to the shape of the separation
distribution, the K.S.\ test prefers to raise $\Md$ to cause a rapid
fall-off below the observed peak.  Doing so sharply reduces the total
lensing probability, however, and thus is disallowed by the $\chi^2$
test.  At this time it is not clear how to interpret these results, or
even how to interpret the sharp peak in the observed
distribution. Given that the sample size is still small and that even
the high bin contains just seven lenses, it could just be due to a
statistical fluctuation.

Nevertheless, the fact that we can reproduce the data so well without
fine-tuning seems to validate our fiducial model.

\section{Forecasts}
\label{sec:Forecasts}

Although the JVAS/CLASS sample is small, ongoing and future surveys
should dramatically increase the size of statistically complete
lens samples and make it possible to constrain the model parameters.
We can use our model to forecast the number of lenses that should
be found in various surveys, and to understand when and how it may
be possible to constrain the various model parameters.  The most
relevant surveys are deep, wide-field optical imaging surveys.  The
sources we consider are quasars, because they are common, distant,
and straightforward to identify on the basis of optical colors
(e.g., Croom \etal 2001; Richards \etal 2002).  See Holz (2001),
Goobar \etal (2002), and Oguri, Suto, \& Turner (2003a) for
estimates of the number of lenses in other populations, such as
supernovae.

To compute the expected number of lensed quasars we must combine
our calculated lensing probability with a description of the quasar
population:
\begin{equation}
  N_{\rm lens} = \int dz \int dS\,\frac{dN_{Q}}{dz\,dS}\,
    P_{\rm lens}(z,S)\ ,
\end{equation}
where $dN_{Q}/dz\,dS$ is the distribution of quasars in redshift and
flux, and $P_{\rm lens}$ is the lensing probability.  We model the
quasar population using the quasar luminosity function derived from
the 2dF quasar redshift survey. Boyle \etal (2000) find that the
luminosity function can be well described as a double power law,
$\Phi \propto [(L/L_*)^\alpha+(L/L_*)^\beta]^{-1}$, where the power
law indices are $\alpha=3.41$ and $\beta=1.58$ and the break
magnitude evolves as $M_*(z) = M_*(0) - 3.4z + 0.68z^2$ for \lcdm.
While the 2dF quasar survey extends to redshift $z=2.3$, we
extrapolate the luminosity function to $z \sim 5$.  Deeper quasar
surveys indicate a bright end slope $\alpha \approx 2.5$ at
$z \sim 4$, suggesting that the pure luminosity evolution model may
break down at high redshift (Fan \etal 2001). This should not
significantly affect our results, however, because even at faint
magnitudes no more than $\sim$10\% of the predicted lenses come
from quasars at $z>4$.  To convert rest frame B-band luminosities
to apparent magnitudes in various observed bands, we compute colors
and $K$-corrections by convolving filter transmission curves with
the composite quasar spectrum given by Vanden Berk \etal (2001).
In the integral to compute the lensing magnification bias (see
eqs.~\ref{eq:bias1} and \ref{eq:bias2}) we use the magnification
of the fainter image, which is appropriate when we want to identify
lenses where {\it both\/} images are above the flux limit.

\begin{figure}
\plotone{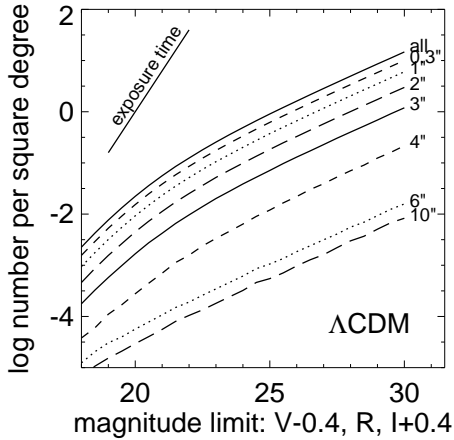}
\caption{
Predicted number counts of lensed quasars as a function of the survey
limiting magnitude, for the fiducial model.  The top curve depicts all
lenses, and the remaining ones show lenses with image separations
$\Dth > 0.3$, 1, 2, 3, 4, 6, 10 arcsec (from top to bottom).  The line
in the upper left corner shows how the survey exposure time scales
with limiting magnitude.
}
\label{fig:NCpred}
\end{figure}

Figure~\ref{fig:NCpred} shows our fiducial predictions for the number
of lensed quasars per square degree, as a function of the R-band
limiting magnitude of a survey, for a variety of cuts on the image
separation.  We have also computed the number counts for the V and I
passbands.  We find that these curves are very similar to those in
Figure~\ref{fig:NCpred} but simply offset horizontally by approximately
$\pm$0.4~mag; thus, we can use a single set of curves to determine
the number counts in any of the three passbands.

An important qualitative result is that the lens number counts
increase less rapidly than the exposure time as the depth increases.
Making a survey 1~mag deeper increases the exposure time by a
factor of 6.3 but only increases the number of lenses by a factor
of $\sim$1.7--3.3 (depending on the original depth).  By contrast,
keeping the same depth but increasing the area by a factor of 6.3
would increase the number of lenses by the same factor of 6.3.  If
the main goal of a survey is to find as many lenses as possible in
a fixed amount of telescope time, then a wide area is more valuable
than a deep limiting magnitude.

\begin{figure}
\plotone{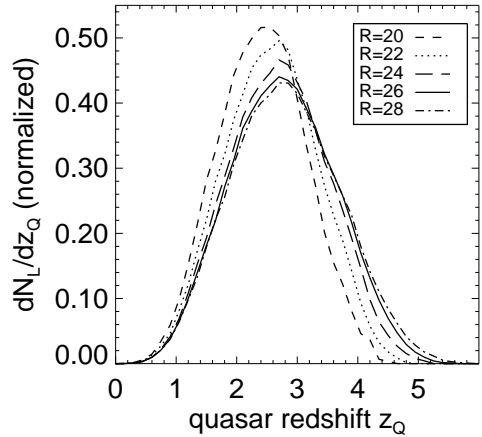}
\caption{
Predicted redshift distribution of lensed quasars.  The different
curves correspond to different R-band limiting magnitudes.
}
\label{fig:Quasardist}
\end{figure}

Figure~\ref{fig:Quasardist} shows the predicted redshift distribution
for lensed quasars.  The distribution peaks at $z \sim 2.5$; it
declines at low redshift mainly because the lensing optical depth
becomes small, and at high redshift because quasars become rare.
The distribution shifts to slightly higher redshifts and becomes
slightly broader at fainter limiting magnitudes, but the change is
not large.  Even at faint limiting magnitudes no more than 10\% of
the predicted lensed quasars have redshifts $z > 4$, so we believe
that uncertainty in the quasar luminosity function at high redshift
(whether the pure luminosity evolution model is viable at
$z \gtrsim 4$) does not significantly affect our results.

Finally, Table~\ref{tab:surveys} gives predictions for the number
of lensed quasars in current and future imaging surveys, obtained
by combining our predicted number counts with the surveys' limiting
magnitudes and areas.  The systematic uncertainties are probably
10--20\% for normal arcsecond-scale lenses, but much larger for
wide-separation lenses (see \S\ref{sec:ParamDep}).  At present about
80 lenses are known, but the largest statistical sample is 13 radio
lenses from CLASS.  The ongoing, ground-based NOAO Deep Wide-Field
Survey (Jannuzi \& Dey 1999) and the Deep Lens Survey\footnote{See
{\tt http://dls.bell-labs.com}.} could each produce a sample as
large as CLASS, or even larger if lenses smaller than 1\arcsec\ can
be identified.  The Sloan Digital Sky Survey by itself should more
than double the number of lenses; several new SDSS lenses have
already been discovered (e.g., Inada \etal 2003a, 2003b; Johnston
\etal 2003).  The proposed Supernova/Acceleration Probe (SNAP; Kim
\etal 2002) satellite should discover a thousand or more lenses,
and measure time delays for the several hundred discovered in the
``deep'' monitoring mode.  The real breakthrough could come from
the Large Synoptic Survey Telescope (Tyson \etal 2002), which would
monitor 3/4 of the sky to reasonable depths and could discover some
10,000 lenses.

\begin{deluxetable}{lclcccc}
\tablecaption{Survey Predictions}
\tablewidth{0pt}
\tablehead{
  \multicolumn{1}{c}{Survey}
 &\multicolumn{1}{c}{Area}
 &\multicolumn{1}{c}{Depth}
 &\multicolumn{4}{c}{Lensed Quasars} \\
 &\multicolumn{1}{c}{($\square\arcdeg$)}
&&\multicolumn{1}{c}{all}
 &\multicolumn{1}{c}{$>\!1\arcsec$}
 &\multicolumn{1}{c}{$>\!3\arcsec$}
 &\multicolumn{1}{c}{$>\!6\arcsec$}
}
\startdata
NOAO DWFS$^{a}$
& 18
& ${\rm R} = 25.8$
& 26
& 11
& 2
& 0.03
\\
DLS$^{b}$
& 28
& ${\rm R} \approx 26$
& 45
& 19
& 4
& 0.1
\\
SDSS$^{c}$
& 10,000
& $i^{*} = 20.2$$^{c}$
& 179
& 73
& 15
& 0.5
\\
SNAP Deep$^{d}$
& 15
& ${\rm I} \sim 30$
& 269
& 110
& 22
& 0.3
\\
SNAP Wide$^{d}$
& 300
& ${\rm I} \sim 28$
& 1,820
& 747
& 147
& 2
\\
LSST$^{e}$
& 30,000
& ${\rm R} \sim 24$
& 14,700
& 6,040
& 1,160
& 18
\\
\enddata
\tablecomments{
$^{a}$NOAO Deep Wide-Field Survey (Jannuzi \& Dey 1999); see
{\tt http://www.noao.edu/noao/noaodeep}.
\\
$^{b}$Deep Lens Survey; see {\tt http://dls.bell-labs.com}.
\\
$^{c}$Sloan Digital Sky Survey (e.g., Abazajian \etal 2003);
see {\tt http://www.sdds.org}.  The SDSS magnitude limit is
$i^{*}=19.1$ for quasars at redshifts $z<3$ and $i^{*}=20.2$ for
quasars at $z>3$ (Richards \etal 2002).
\\
$^{d}$Supernova/Acceleration Probe; see {\tt http://snap.lbl.gov}.
Current mission parameters include a ``deep'' mode with deep
monitoring of 15~$\square\arcdeg$, and a ``wide'' mode with shallower
imaging of 300~$\square\arcdeg$ (Kim \etal 2002).
\\
$^{e}$Large-aperture Synoptic Survey Telescope (Tyson \etal 2002);
see {\tt http://www.lsst.org}.
}
\label{tab:surveys}
\end{deluxetable}

The search for lenses in these surveys will exploit the multiple
colors that they all contain.  Color cuts can be used to identify
quasar candidates (e.g., Croom \etal 2001; Richards \etal 2002).
Lens candidates can then be selected either as pairs of quasars
separated by a few arcseconds, or as objects that have quasar
colors but are marginally resolved (i.e., small-separation lenses
in ground-based surveys).  Such lens search techniques are being
refined for large surveys by SDSS (Inada \etal 2003a; Pindor \etal
2003).

Spectroscopic confirmation of faint lens candidates may be difficult
at present, although not with future 30-meter class telescope.  Still,
other confirmation methods are available.  First, candidates with four
(or more) lensed images can often be confirmed by the image
configurations alone, and this should account for at least 25--30\%
and perhaps as many as 60\% of the lenses (based on current predicted
and observed 4-image to 2-image lens ratios; Keeton, Kochanek, \&
Seljak 1997; Rusin \& Tegmark 2001).  Second, surveys that include
near-infrared imaging may discover infrared Einstein rings, which are
lensed images of the quasar host galaxies that provide unambiguous
evidence for lensing (e.g., Kochanek, Keeton, \& McLeod 2001).
Finally, monitoring programs like SNAP and LSST would offer the first
chance to confirm lenses via time delays.

\section{Conclusions}
\label{sec:Conclusions}

The statistics of strong gravitational lenses depend on a number
of parameters related to both cosmology and the internal structure
of the lenses.  We have developed the full formalism for computing
lens statistics in universes dominated by quintessence.  We have
also shown how variations in the parameters lead to changes in the
distribution of image separations, time delays, and lens redshifts.
The strongest effects are found in the image separation distribution.
Most lenses have image separations of around one arcsecond,
corresponding to lensing by individual galaxies.  In this regime,
the most important parameters are $w$ and $\s_8$, which produce
changes in the number of lenses at the tens of percent level.
These effects are too small to be probed with existing lens surveys
like CLASS (but see Huterer \& Ma 2003), but will not be out of
reach of the large samples that will be found with ongoing and
future surveys.

The most dramatic parameter dependences are found among wide-separation
lenses produced by groups and clusters of galaxies.  The abundance
of these lenses is extremely sensitive to the distribution of
``concentration'' parameters that describe the inner structure of
massive dark matter halos, and somewhat less sensitive to the
abundance of such halos.  Wide-separation lenses will always be
rare, but the remarkable sensitivity to the parameters means that
even a small number of lenses with separations greater than
$\sim\!5\arcsec$ should yield interesting constraints.  Indeed,
the first instance of a wide-separation lens in a statistical sample
has recently been discovered and is being analyzed for constraints
on $\s_8$ and the core structure of cluster halos (Inada \etal 2003b;
Oguri \etal 2003b).

Imaging surveys that are already underway will substantially
increase the sample of strong lenses, and future surveys promise even
more.  Predicted samples of hundreds or thousands of lenses will
revolutionize lens statistics, provided that their selection effects
are well understood.  A robust measurement of the distribution of
image separations will yield internally self-consistent tests of
cosmological parameters, the dark matter density profiles predicted
by the popular cold dark matter, and the physics of baryonic cooling
in massive halos.

\acknowledgements 
We thank J.~Frieman, M.~Oguri, D.~Rusin, and L.~Strigari for
interesting and helpful discussions.  Support for this work was
provided by NSF grant AST-0205738 (P.M.).  and by NASA through Hubble
Fellowship grant HST-HF-01141.01-A (C.R.K.)  from the Space Telescope
Science Institute, which is operated by the Association of
Universities for Research in Astronomy, Inc., under NASA contract
NAS5-26555.

\appendix

\section{Power Spectrum Normalization}

We have normalized our power spectrum to match the observed abundance
of galaxy clusters today. The largest and most recent X-ray flux-limited 
sample has been assembled by Reiprich \& B\"ohringer (2002) and is known
as HIFLUGCS. The cluster sample is based on the ROSAT All-Sky Survey
and is composed of the bightest 63 clusters with galactic latitude
$|b_{II}|\geq20^{\circ}$ and flux
$f_X(0.1-2.4 \mbox{ keV})\geq2 \times 10^{-11}\,$ ergs s$^{-1}$ cm$^{-2}$.

From the published data (Reiprich \& B\"ohringer 2002) we have taken
the cluster redshift ($z$), the flux in the energy range $0.1-2.4$ keV
($f_X$), and $M_{200}$, the mass enclosed in a region in which the average
mass density is 200 times the critical density, and calculated an X-ray
cluster mass function, analogous to the work by Reiprich \& B\"ohringer
(2002). The mass function was determined using the classical $V_\max$
method:
\be
\f{dn}{dM}=\f{dN}{dVdM}=\f{1}{\De M} \sum_{i=1}^N\f{1}{V_{\max,i}},
\ee
where $\De M$ denotes the width of the mass bin, and $V_{\max,i}=V_\max 
(L_{X,i})$ is the maximum comoving volume within which the
$i^{\mbox{\small th}}$ cluster with luminosity $L_{X,i}$ could have
been detected given the flux limit of the sample. Reiprich \& B\"ohringer
(2002) found a correlation between the total mass of the clusters and
their X-ray luminosity, but with a significant scatter.  An advantage 
of using  $V_\max(L_X)$ in the calculation of the mass function is 
that this scatter is automatically taken into account. 

There is a slight cosmology dependence of the mass function calculated
this way, stemming from the use of the luminosity distance in converting
$f_X$ to $L_X$, but the effect is never larger than $5\%$ and always
smaller than the Poisson noise due to the limited number of clusters
per bin. 

\begin{figure}
\includegraphics[width=3.1in]{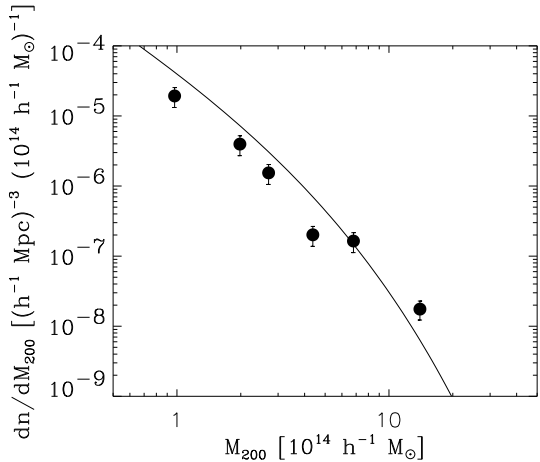}
\caption{\lcdm HIFLUGCS data and best-fit ($\s_8=0.74$)
Jenkins \etal (2001) mass function.}
\label{fig:massfunction}
\end{figure}

The last step in the cluster normalization of the power spectrum is the 
matching of a theoretical mass function. As explained in \S~\ref{sec:massf}
we have chosen the Jenkins \etal (2001) mass function, which depends on
the power spectrum through $\sigma_M$, see eqs.~(\ref{eq:var}) and
(\ref{eq:massfunctions}).  We simply performed a $\chi^2$ minimization
to find the value of $\s_8$ that produced the best-fit between the
mass function and the HIFLUGCS X-ray cluster mass function. The $w=-1$
mass function and the best-fit mass function are shown in
Figure~\ref{fig:massfunction}. The $w$-dependence enters this method in
two minor ways: the first is the aforementioned cosmology dependence
of $V_\max$, and the second arises from the transfer function.  Neither
of these two effects is very significant, however, and so it is no
surprise that the resulting $\s_8$'s are identical within 
uncertainties.  We find $\s_8=0.738, 0.740,$ and 0.742 for $w=-1,
-0.75, \mbox{and} -0.5$, respectively.  The value $\s_8=0.74$ is 
consistent with what Reiprich \& B\"ohringer (2002) found, who performed
a more extensive analysis and found $\s_8=0.43 \Omega_M^{-0.38}$,
which evaluates to $\s_8=0.68$ for $\Omega_M=0.3$.  Note, however,
that their full analysis prefers $\Omega_M=0.12^{+0.06}_{-0.04}$ and 
$\s_8=0.96^{+0.15}_{-0.12}$.

{}

\end{document}